# Sub-20 nm Nanopores Sculptured by a Single Nanosecond Laser Pulse

Yanwen Yuan,[*] Guangyuan Li,[†] Hamed Zaribafzadeh,[†] María de la Mata,[¶] Pablo Castro-Hartmann,[‡] Qing Zhang,[†] Jordi Arbiol,[¶,§] and Qihua Xiong,[†,#]

[*] Centre for Advanced 2D Materials, National University of Singapore 6 Science Drive 2, Singapore 117546

[†] Division of Physics and Applied Physics, School of Physical and Mathematical Sciences, Nanyang Technological University, Singapore 637371.

[¶] Institut de Ciència de Materials de Barcelona, ICMAB-CSIC, Campus de la UAB, 08193 Bellaterra, Catalonia, Spain.

[‡] Servei de Microscòpia, Universitat Autònoma de Barcelona (UAB), E-08193 Bellaterra, CAT, Spain

[§] Institució Catalana de Recerca i Estudis Avançats (ICREA), 08010 Barcelona, Catalonia, Spain.

[#] NOVITAS, Nanoelectronics Centre of Excellence, School of Electrical and Electronic Engineering, Nanyang Technological University, Singapore, 639798.

**ABSTRACT**

Single-molecule based 3$^{rd}$ generation DNA sequencing technologies have been explored with tremendous effort, among which nanopore sequencing is considered as one of the most promising to achieve the goal of $1,000 genome project towards personalized medicine. Solid state nanopore is consented to be complementary to protein nanopore and subjected to extensive investigations in the past decade. However, the prevailing solid-state nanopore preparation still relies on focused ion or electron beams, which are expensive and time consuming. Here we demonstrate the fabrication of nanopores down to 19 nm with a single nanosecond laser pulse. The laser drilling process is understood based upon a 2D axisymmetric transient heat transfer model, which predicts the laser fluence-dependent pore size distribution and shape with excellent agreement to electron microscopy and tomography analysis. As-drilled nanopore devices (26 nm) exhibit adequate sensitivity to detect single DNA molecule translocations and discriminate unfolded or folded events. Sub-10 nm nanopores can be readily achieved upon a thin layer of alumina deposition by atomic layer deposition, which further improves the DNA translocation signal to noise ratio considerably. Our work provides a solution for fast, low-cost and efficient large-scale fabrication of solid state nanopore devices for the 3$^{rd}$ generation nanopore sequencing.

Nanopore sequencing was coined in a pioneering work of using an α-hemolysin protein ion channel imbedded in a lipid bilayer to detect single DNA molecules.[1-3] With two buffer-filled reservoirs connected only by the nano-size channel (or called nanopore), single-stranded DNA molecules electrophoretically translocate through the nanopore upon a voltage bias supplied, leading to a series of ionic current blockade events.[4] Depending on the physical sizes of the individual bases on a single-stranded DNA chain, the sequencing of DNA could be read directly from the unique blockade current signal with a high speed regardless of the DNA length.[5-7] However, protein nanopore has a well-defined size of ~ 1.5 nm, exhibits no freedom to tune the pore size. In addition, it also suffers from instability as being labile. To develop a more flexible and stable platform for nanopore sensing and sequencing, solid-state nanopore across insulating solid-state membrane demonstrate its ability to sense a variety of bio-molecules including DNA, RNA and proteins.[8-12] Based on the extensive range of materials and size request, chemical etching,[13] ion or electron beam sculpting[14,15] have been explored in nanopore preparation, among which focused electron beam drilling using a transmission electron microscope (TEM) is the most widely used approach because of high precision and less materials limitation.[16-18] Nonetheless, TEM drilling is time-consuming and expensive to operate as usually users have to align the beam carefully and fight against the thermal drift of sample, making the nanopore device research accessible to only handful laboratories worldwide.

Lasers are widely used in drilling and cutting of materials for industrial manufacturing applications,[19] in which micro-size drilling in various materials can be prepared efficiently and economically without contamination.[20] By use of high numerical aperture optics, short wavelength light sources (*e.g.*, ultra violet) and femtosecond laser duration, sharp and well-defined nanostructures of a few hundred nanometers can be produced by laser manufacturing.[21-22] With enhanced femtosecond pulse laser such as near field laser drilling and liquid-assisted lasers drilling, the resolution can be further improved to sub-100 nm.[23-25] However, direct laser drilling still faces considerable challenges to achieve nano-scale (<100 nm) resolution constricted by the diffraction limit.[26] Surprisingly, we present in this paper for the first time that nanopores down to 19 nm can be fabricated with a single nanosecond laser pulse across a SiNx membrane. The size is not determined by diffraction limit, but by the kinetic process dependent by the laser fluence and pulse duration. The drilling-through time is only within a few nanoseconds, thermal drift of the sample is no longer a concern in the fabrication.

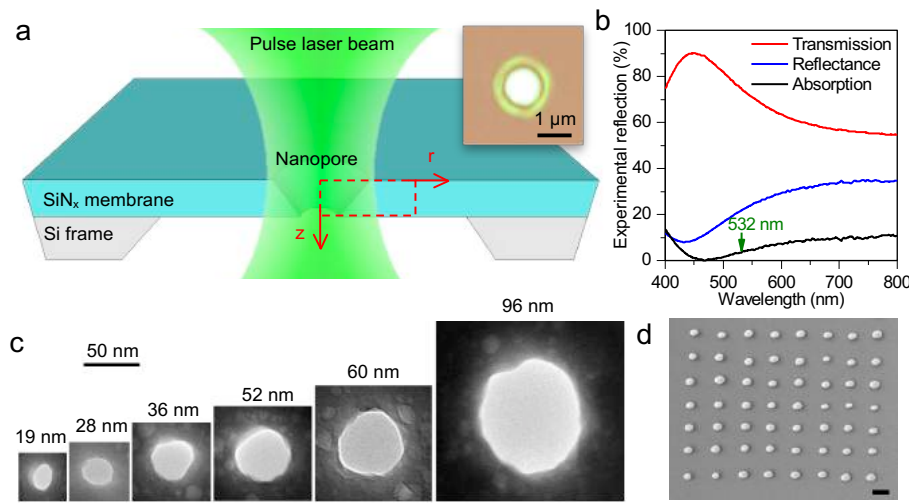

**Figure 1.** Schematic of laser drilling and laser drilled nanopores. (a) Schematic of laser drilling: A focused 532 nm Gaussian profile pulsed laser (inset) impinges onto the SiNx membrane window supported by a silicon frame. The red dashed box illustrates the simulation domain, and the inset shows the focused laser beam. (b) The measured transmission, reflection and the deduced absorption spectra of the 100 nm thick SiNx membrane. (c) The TEM images of drilled nanopores with diameters ranging from 19 nm to 96 nm. (d) A 7· 8 arrayed laser drilled nanopores with the scale bar of 1 μm.

A 532 nm Nd:YAG nanosecond pulsed laser is focused into a Gaussian beam profile with a spot size of 1 μm onto the silicon nitride (SiN$_x$) membrane surface (schematic diagram shown in Figure 1a). About 3.6% of the laser power is absorbed as measured by a microspectrophotometer (Figure 1b) and presumably converted into heat, which results in the vaporization of SiN$_x$ and the drilling of a pore in the membrane. By varying the laser power, the nanopore diameter can be tuned from several hundred to tens of nanometers (Figure 1c). The smallest nanopore we achieved using manual adjustment is ~ 19 nm in diameter. With this rapid and efficient process, arrayed nanopores of desired patterns can be realized readily even with a manual translation of the microscope stage (Figure 1d).

To understand the underlying laser drilling mechanism, we establish a microscopic theoretical model to describe the laser drilling process in the red dashed area (Figure 1a). The amorphous low-stress SiN$_x$ is considered to be isotropic, and laser induced heat source is assumed to be distributed in a Gaussian profile along the radial direction ($r$) and to be constant during the pulse-on time for a single pulse. The energy balance at the top surface is expressed as[27]

$$\mathbf{n} \cdot (-k \nabla T) = 2\alpha_{\text{abs}} P / (\pi R^2) \exp(-2r^2/R^2) \qquad (1)$$

where $\mathbf{n}$ is the normal vector of the boundary, $k$ is the thermal conductivity of the SiN$_x$ film, $T$ is the temperature, $\alpha_{\text{abs}}$ is the membrane absorption coefficient, $P$ is the laser power, and $R$ is the radius of focused laser beam. The left hand side describes the heat flux along the surface normal, while the right hand side corresponds to the boundary heat source, which is induced by the absorbed laser power averaged over the laser spot. Since the laser beam diameter is much smaller than the SiN$_x$ membrane lateral size, a 2D axisymmetric transient heat transfer model is used assuming that the laser energy absorbed at the absorption front converts to heat instantaneously (method). Because the thermal properties, such as the thermal conductivity, the density and the specific heat differ after the vaporization of SiN$_x$, we define a parameter, $H(T)$, which varies from zero for solid to be unity for vapor, to track the phase change process. As the vaporized material leaves the surface and then the laser marches forward to strike the newly exposed absorption front, there should be a dynamic computation domain, meshing, and boundary heat source modeling, making the computation very complex and thus time-consuming. Instead, we tackle these effects purely by using a mathematical approach,[28] which is simple to implement numerically. A static computation domain, meshing and boundary heat source modeling is used. We set the density and specific heat for a location, where the phase parameter $H$ reaches unity, to be that of air to remove the amount of thermal energy carried away by the vapor jet. To virtually apply the laser heat source to the newly exposed surface, which predominantly evolves in the beam propagation direction ($z$), we elevate the thermal conductivity in $z$ direction to twelve times of that of the material, and meanwhile reduce the thermal conductivity in $r$ direction to two thirds of that of the material.

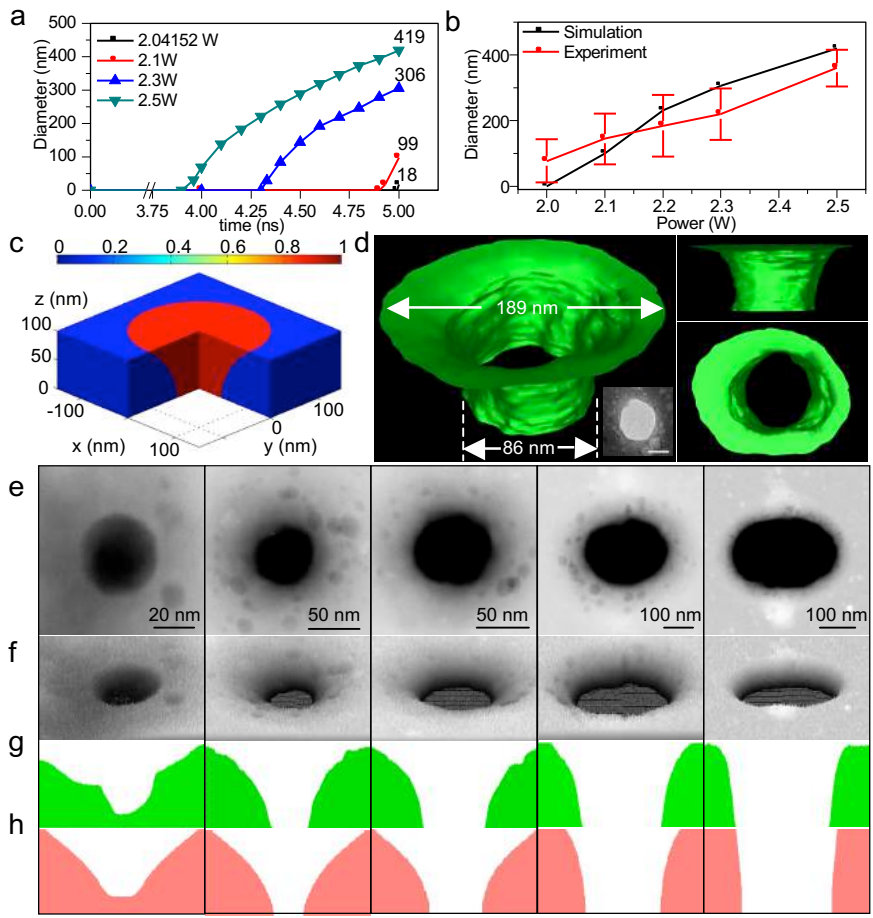

**Figure 2.** Experimental and simulated tomography of laser drilled nanopores. (a) Simulated nanopore size evolution *versus* time under various laser power from 2.04152 W to 2.5 W. (b) Experimental and simulated nanopore sizes *versus* the laser power. (c) Simulated nanopore of 198 nm on the top and of 87 nm on the bottom (power: 2.05 W). (d) Left: The side view and top view of the reconstructed model; Right: The STEM reconstructed image of a 86 nm drilled nanopore (inset scale bar is 50 nm) which shows 189 nm diameter on the top surface and 86 nm on the bottom. (e-g) HAADF STEM top view, tilted view and the reconstructed cross section images of drilled nanopores of non-through hole, 49 nm, 70 nm, 120 nm, 200 nm and 265 nm (from left to right). (h) The simulated cross section profiles of drilled nanopores of non-through hole, 51 nm, 66 nm, 123 nm, 183 nm and 246 nm (from left to right).

Figure 2 summarizes the computational results and the electron microscopy characterizations of the pores. As shown in Figure 2a, our model predicts that: (1) there exist a critical fluence (2.04152 W), smaller than that it is impossible to drill a pore; (2) above the threshold fluence a fairly long "incubation" time is needed within the pulse duration, the higher the laser fluence, the shorter the incubation time; (3) once the pore is drilled through, it continues to grow within a single pulse until reaching the end of 5 ns, giving rise to the ultimate pore size we observed. For the critical fluence, the simulated pore diameter is only 18 nm. This is very close to what was achieved in experiment (19 nm). Moreover, the simulated pore sizes for various laser powers show good agreement with the experimental average from accurate electron microscopy measurement (Figure 2b). The size deviations in experiment may originate from the instability of the laser power (~ 10%) and the duration time. Such deep subwavelength nanopore sizes are due to the fact that Gaussian-shaped laser power flows predominantly in *z* direction, as we have stated in the model. As such, our model predicts that the actual pore exhibits a funnel-like shape, with the top surface being the wider opening, while the bottom part is more or less cylindrical (shown in Figure 2c). To confirm the prediction, we performed electron tomography of reconstruction of a nanopore based upon a tilt series of bright field (BF) transmission electron microscopy (TEM) images. Figure 2d shows that the reconstructed 3D structure of a laser drilled nanopore is in a homogeneous shape with a funnel-like morphology, which is in good agreement with the profile of the simulation model (Figures 2c). Based on high angle annular dark filed (HAADF) scanning TEM (STEM) images obtained in several nanopores with different sizes (Figure 2e-g) we obtained the corresponding surface plots and profiles, which agree well with the electron tomography reconstruction and simulation results (Figure 2h). Both simulated and experimental results show that the morphology holds funnel shape from non-through pores to small ones, and then become more cylindrical for large diameters, validating that the vaporization process is mainly in *z* direction. Two video clips are provided in the supplementary information to demonstrate the simulated laser drilling process (and the drill-through shape) and the electron tomography demonstration of the nanopore shape presented in Figure 2d.

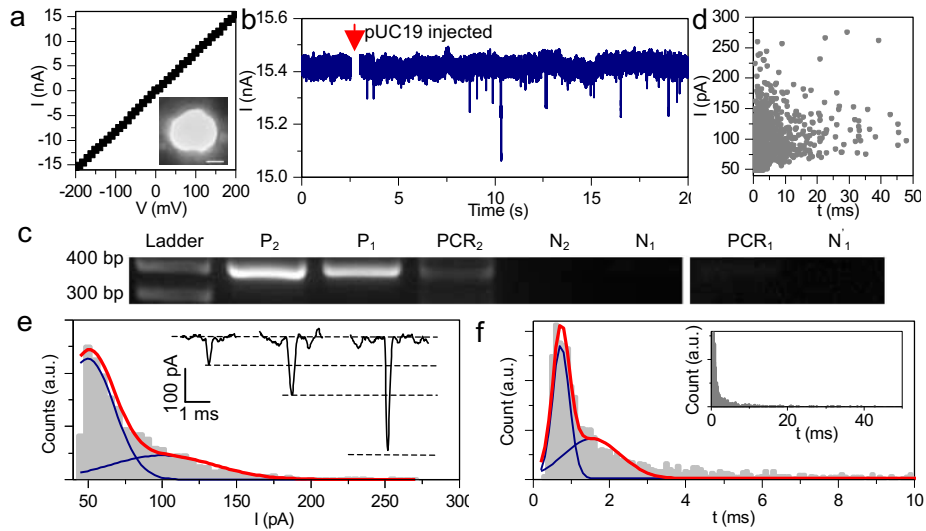

**Figure 3.** DNA translocation using laser drilled nanopore. (a) I-V characterization of a 26 nm laser drilled nanopore (shown in the inset with the scale bar being 10 nm) in 1 M KCl. (b) DNA translocation events through the nanopore under a bias of 200 mV. (c) PCR result of DNA molecule in *trans* chamber after translocation, in which ladder is the DNA ladder to label our 400 bp target DNA designed for pUC19, $P_1$ and $N_1$ is the positive and negative controls for the first PCR, and $P_2$, $N_2$ for the second PCR. PCR2 shows the result of second PCR while $PCR_1$ and $N'_1$ in the second column is the result of the first PCR. (d) Scatter diagram of translocation events (*N*=1889) by blockade current and dwell time. (e) Histogram of counts versus blockade current. The insert shows three typical translocation events for unfolded, folded and double-folded DNA translocations. (f) Histogram of counts *versus* dwell time from 0 to 10 ms with an inset showing the whole view.

Laser drilled nanopore with its sub-100 nm size suggested its capability for sensing of single molecules including DNA. We demonstrate the nanopore single molecule sensing of a 2686 bp linearized double-stranded DNA, namely pUC19[29] (method). Figure 3a displays a linear I-V characterization of a 26 nm SiNx pore with a thickness of 100 nm. With both chambers containing 1.0 M KCl buffer solutions, the device shows a conductance of 78.7 nS. After pUC19 molecules injected into *cis* chamber and a 200 mV bias applied across the nanopore, ionic current through the nanopore have been recorded as plotted in Figure 3b. A baseline current of ~ 15.4 nA and a noise level of about 24 pA have been observed. A series of current blockade events are recorded with different blockade current magnitude. By reversing the bias, the DNA translocation event were observed to cease (data not shown here).

To confirm that the translocation event observed indeed correspond to pUC19 molecules and exclude the possibility of random telegraph noise,[30] we performed the polymerase chain reaction (PCR) to indentify the DNA molecules in *cis* and *trans* chambers (methods). We use the DNA injected buffer in *cis* chamber as the positive control ($P_1$, $P_2$), and distilled water is used as the negative control ($N_1$, $N_2$, $N'_1$) in order to compare with the PCR result in *trans* chamber after DNA translocation ($PCR_1$, $PCR_2$). Both positive control shows bright bands in the expected area (400 pb in ladder), and negative controls are totally blank (Figure 3c). The single round PCR ($PCR_1$) products shows a very poor fluorescence mark due to the low concentration of DNA molecules after translocation. To improve the sensitivity, another PCR is done with two rounds of PCR process ($PCR_2$) and shows a clear white band in the labeled region, indicating that translocations of pUC19 molecule dominate the current blockades trace.

The recorded translocation events exhibit a large span of blockade magnitude and dwell time which indicate diverse translocation modes in the nanopore device as shown in the scatter plot (Figure 3d). We also analyze the histogram of the blockade current magnitude shown in Figure 3e. The histogram revealed two most probable peaks, with the current magnitude around ~ 65 and ~ 130 pA, which is likely corresponding to unfolded and folded DNA translocation events, respectively. Since the nanopore diameter is 10 times bigger than DNA diameter, even a few double folded DNA translocations (~ 260 pA) has been recorded as well, which however is statistically not significant to show in the histogram. Furthermore, the dwell time of unfolded DNA (0.75 ms) is half of that of the folded ones (1.5 ms) (Figure 3f).

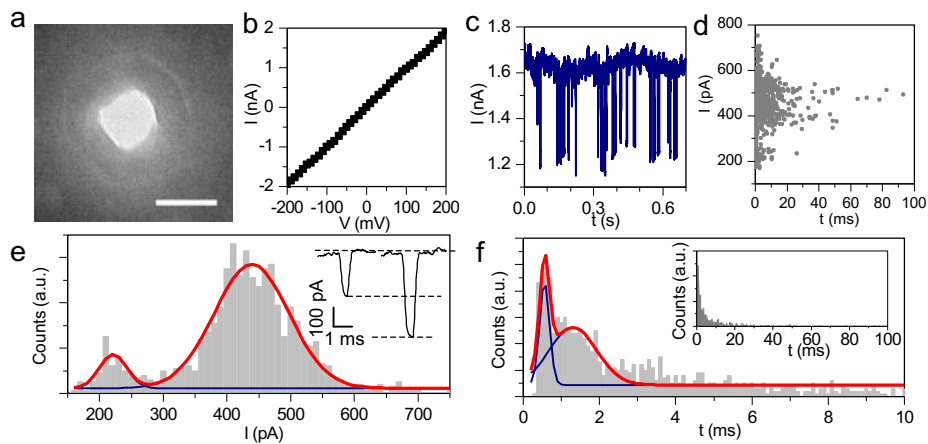

**Figure 4.** DNA translocation using ALD shrunk nanopore. (a) A TEM image of a 10 nm shrunk nanopore after 5 nm $Al_2O_3$ ALD (scale bar of 10 nm). (b) I-V characterization of this nanopore in 1 M KCl. (c) DNA translocation events through the ALD shrunk nanopore under a bias of 200 mV. (d) Scatter diagram of translocation events (*N*=1504) by blockade current and dwell time. (e) Histogram of counts versus blockade current. The inset shows two typical translocation events for unfolded and folded DNA translocation. (f) Histogram of counts *versus* dwell time from 0 to 10 ms with an inset showing the whole view.

The smallest size we can achieve at present is still larger than the diameter of DNA molecules. To put it in perspective, if the laser pulse width and the power density can be accurately controlled, one should be able to reproducibly drill nanopores within 10 nm. Another option is to add one more step of atomic layer deposition, which has been shown to effectively decrease the size and change the surface functionality of nanopore. By coating a 5 nm alumina using ALD, we are able to shrink a 20 nm nanopore to 10 nm (Figure 4a). The corresponding conductance of a 10 nm ALD shrunk nanopore is decreased to 10 nS (Figure 4b), and the baseline ionic current has been decreased dramatically (from 15 nA to 1.5 nA) with the noise level ~ 17 pA. Under the same sensing condition, a series of sharp translocations of pUC19 molecules have been achieved (Figure 4c). The ALD shrunk nanopore with a smaller diameter has larger electric field intensity around, as a result, the typical blockade currents for the unfolded and folded events increase to ~ 220 pA and ~ 440 pA, and the signal to noise ratio (SNR) is greatly improved compare with the as-drilled nanopore. Furthermore, because of a much bigger resistance in a small and long ALD-shrunk nanopore, folded events require a longer dwell time (0.6 ms) compared with unfolded ones (1.4 ms) during the translocation (Figure 4f).

In summary, we have proposed and realized a simple and rapid fabrication of sub-20 nm solid-state nanopores in $SiN_x$ membrane using a single nanosecond laser pulse, far beyond the optical diffraction limit. The laser drilling can be well understood based upon a 2D axisymmetric heat transfer model. We expect this model is applicable in other materials and exhibit the potential to push further down to sub-10 nm. Our laser drilling method provides a new commercial nanopore fabrication approach and pushes forward the frontier of fast and cheap nanopores for high-sensitive single molecule detection.

**METHODS**

**Laser drilling and characterization of nanopore**

A nanosecond solid state laser (Quantel Brilliant B, 1064 nm) is used to produce a nanosecond-pulse laser beam with a temporal width of about 5 ns, 532 nm wavelength (using a frequency doubler) and 10 Hz repetition rate. The controller is able to release only one pulse. The pulsed laser beam is focused onto a 0.1· 0.1 mm$^2$, 100 nm thick SiN$_x$ membrane window (SPI) by an objective (100X, numerical aperture = 0.9).

The images of laser drilled pore are taken by TEM (JEOL 1400). For tomographic analysis, single-tilt series (JEOL EM-21010 rod with EM-21020 retainer) were collected at 60,000-fold magnifications (pixel size of 0.33 nm) using a JEOL JEM 2011 operated at 200 kV coupled to a 2048 × 2048 Gatan Ultrascan 1000 CCD camera. The 3D Tomography-Acquisition Software package (Gatan) was used to acquire tilt series, from −42° to +41°, with a 1° increment. The IMOD software package was used for the entire procedure of image alignment and reconstruction.[31]

High angle annular dark field (HAADF) scanning transmission electron microscopy (STEM) images were obtained in different nanopore sizes. The HAADF STEM measurements were performed in a FEI F20 field emission gun microscope operated at 200 kV. As the HAADF STEM intensity collected in the images is barely proportional to Z$^2$ and the sample thickness, we could obtain 3D surface plots on different nanopores showing detailed information about their morphology (in good agreement with the e-tomography reconstructions). This procedure had been previously successfully used in order to obtain accurate 3D morphology reconstructions in non-planar nanostructures such as nanowires.[32-33]

**Modeling and Simulation**

We use an axisymmetry boundary condition at $r = 0$, and an open boundary condition at $r = 3a$ since the radial dimension of the membrane is much larger than the laser diameter. The governing equation for the transient heat transfer process is expressed as follows,[27, 34]

$$\rho C_p \partial T / \partial t - \nabla \cdot (k \nabla T) = 0 \qquad (2)$$

where $\rho$ is the density and $C_p$ is the specific heat. On the bottom surface of the membrane ($z = 0$), convection and surface-to-ambient radiation cool the system. The ambient and the initial conditions of the domain are assumed to be of room temperature. This 2D surface-heating source model can explain funnel-shaped profiles in pulsed laser machining in both opaque and transparent materials.[35-36] We have also simulated using a bulk-heating model which predicted an asymmetric truncated double-cone shape as similarly reported in nanopores fabricated by electron beam drilling.[37] This is contradictory to the experimental nanopore profile observed in our experiments, therefore the bulk-heating model was not adopted in our simulation.

During the vaporization process, which starts at $T = T_1$ and end at $T = T_2$, the latent heat is incorporated by modifying the specific heat,

$$C_p(T) = \begin{cases} C_{p,s} & \text{for } T < T_1 \\ C_{p,s} + D(T)l_m & \text{for } T_1 < T < T_2 \\ C_{p,\text{air}} & \text{for } T > T_2 \end{cases} \qquad (3)$$

where $C_{p,s}$ is the specific heat of solid $SiN_x$, $C_{p,\text{air}}$ is the specific heat of air (or vapor $SiN_x$), $D(T) = dH(T)/dT$, and $l_m$ is the latent heat of fusion. The model is numerically calculated using the finite element method in the commercial multiphysics software-COMSOL. An animation of the simulated laser drilling process is provided in supplementary information, which illustrates how the nanopore evolves upon laser irradiation to form a funnel-shape nanopore.

**DNA translocation measurement**

The nanopore chip was glued to a PVC chip holder by Epoxy resins and two PVC tubes are also glued to the two sides of chip holder to form the *cis* and *trans* chambers. The device is mounted in a Faraday box to avoid outside electromagnetic interference. Axopatch 200B capacitive feedback patch clamp is used to apply the bias and detect the ionic current signals through a pair of Ag/AgCl electrodes. The current signal is recorded with a 50 kHz sampling rate and filtered by 2 kHz Bessel filter, then further digitized using Axon Digidata 1440A data acquisition system.

**DNA handling and PCR**

The flat-end linearized pUC19 was prepared from circular vector pUC19 after restriction enzyme digestion by Sma1 for 1 h in room temperature, and purified by a PCR purification kit (QIAGEN GmbH, Germany). 1 nM linearized pUC19 is used in DNA translocation experiments.

For PCR amplification, the translocated DNA molecules are filtered by the PCR purification kit and amplified for 35 times. The PCR reaction mixture shown in Figure 3d was prepared by mixing forward and reverse primers (designed as 5'GATCCGGCAAACAAACCACC3' and 5'TGGGTCTCGCGGTATCATTG3' respectively, AITbiotech Singapore), dNTPs (Thermo Fisher Scientific, USA), $MgCl_2$, GoTaq Flexi Buffer, and GoTaq DNA Polymerase (Promega, USA). PCR was optimized for pUC19 as 1 min initial denaturation at 94°C, 35 cycles of 30 sec denaturation at 94°C, 30 sec annealing at 65°C, and 1 min extension at 72 °C, and 5 min final extension at 72°C. *Cis* and *trans* chamber solutions were separately taken for PCR. Since the number of molecules in *trans* chamber after translocation is very low, it is not possible to produce proper bands by regular PCR protocol.[38-39] Therefore to concentrate the molecules, *trans* chamber solution was first desalted by the kit, frozen overnight at -20°C, and finally freeze dried (Freeze dryer ALPHA 1-2 LDplus, Martin Christ Gefriertrocknungsanlagen GmbH, Germany) and dissolved into PCR reaction mixture to perform PCR. To further amplify PCR band signal, *trans* chamber PCR product was again desalted, freeze dried, and dissolved into PCR reaction mixture and PCR was repeated. In both PCRs, *cis* chamber solution was used for positive control without any purification.

## ASSOCIATED CONTENT

**Supporting Information**

An animation of the simulated laser drilling process and a video clip of laser drilled nanopore e-tomography reconstruction are provided in supplementary information.

## AUTHOR INFORMATION

**Corresponding Author**

*E-mail: C2DYY@nus.edu.sg

**Author contributions**

Y.Y. and Q.X. conceived the idea, Y.Y. and Q.Z. devised and performed the laser drilling, G.L. performed the modeling and simulation, Y.Y. conducted the DNA translocation experiments and analyzed the data, H.Z. performed the PCR experiments, M.d.l.M., P.C.H. and J.A. performed the e-tomography and HAADF STEM surface plot analyses, Y.Y., G.L. and Q.X. wrote the manuscript, all the authors commented on the manuscript.

## ACKNOWLEDGMENT

Y.Y. and Q.X. acknowledges the support from the Singapore National Research Foundation through a fellowship grant (NRF-RF2009-06). This work was also supported in part by the SPMS collaborative fund (M4080535) and start-up grant support (M58113004) from Nanyang

Technological University. Q.X. gratefully thanks Nanyang Nanofabrication Center of for the help on the ALD deposition. J.A. acknowledges the funding from the Spanish MICINN project MAT2010-15138 (COPEON) and Generalitat de Catalunya (2009 SGR 770 and NanoAraCat). M.d.l.M. thanks CSIC JAE Pre-Doc program. We acknowledge the use of STEM facilities at ICN2 and UAB, as well as Dr. Belen Ballesteros for helpful assistance.

Table of Contents

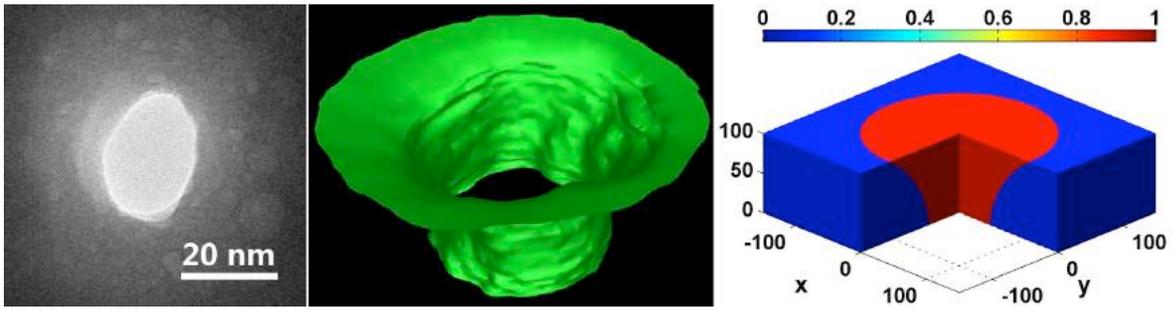